\documentclass[12pt]{amsart}
\usepackage{geometry}            
\usepackage{graphicx}
\usepackage{amssymb}
\usepackage{mathtools}
\usepackage{epstopdf}
\usepackage{color}
\usepackage{mathpazo}
\fontfamily{Palatino}
\newcommand{\be}{\begin{equation}}
\newcommand{\ee}{\end{equation}}
\newcommand{\ba}{\begin{eqnarray}}
\newcommand{\ea}{\end{eqnarray}}

\begin{document}

\begin{center}
\tiny{Essay written for the Gravity Research Foundation 2022
Awards for Essays on Gravitation}

\vspace{1.5cm}
\huge{\bf Emergent Early Universe Cosmology from BFSS Matrix Theory}
\vspace{1cm}

\large{\bf  Suddhasattwa Brahma$^*$, Robert Brandenberger$^\dagger$ and\\ Samuel Laliberte$^\dagger$}  \\
\vspace{1cm}
\end{center}
{\tiny{
$^*$Higgs Centre for Theoretical Physics, School of Physics \& Astronomy,
University of Edinburgh, Edinburgh EH9 3FD, UK.  \\
$^\dagger$Department of Physics, McGill University, Montr\'{e}al, QC, H3A 2T8, Canada.\\
\vspace{1cm}
\begin{center}
{\large{\bf Abstract}}
\end{center}
\small{The BFSS matrix model is a suggested non-perturbative definition of string theory. Starting from a thermal state of this matrix model, we show how space and time can emerge dynamically. Results from the IKKT matrix model indicate that the $SO(9)$ symmetry of space is spontaneously broken to $SO(3) \times SO(6)$, with the three-dimensional subspace becoming large. Given this initial state for the universe,  we show that cosmological perturbations and gravitational waves with scale-invariant spectra are generated, without the need of postulating an early phase of cosmological inflation. The Big Bang singularity is automatically resolved.}

\vspace{1.5cm}
\noindent email for SB: sbrahma@ed.ac.uk \\
email for RB: rhb@physics.mcgill.ca \\
email for SL: samuel.laliberte@mail.mcgill.ca \\
\\
Corresponding author: Robert Brandenberger\\
Submission date: March 11 2022

\date{\today}                                           


\newpage
\normalsize
 
Superstring theory is a promising candidate for a unified quantum theory of space, time and matter. General arguments \cite{GSW} indicate that the theory should be able to resolve the singularities and divergences which plague General Relativity on one hand, and derive quantum field theories underlying the standard model of matter on the other. However, string theory is usally considered at a perturbative level and at the level of an effective field theory on a classical background space-time. However, there is mounting evidence that such a description will break down at very high densities and curvatures, in particular in the very early universe \cite{swamp, TCC}.

A number of years ago, it was suggested that in the large $N$ limit, a quantum mechanical model of $N \times N$ Hermitean matrices might provide a non-perturbative definition of string theory \cite{BFSS}. The action for this model is given by \footnote{Note that the indices are contracted using the Minkowski $\eta_{\mu \nu}$ tensor.}
\be \label{BFSSaction}
S \, = \, \frac{1}{2 g^2} \int dt \bigl[ {\rm Tr} \left \{\frac{1}{2} (D_t X_i)^2 - \frac{1}{4} [X_i, X_j]^2\right\}
 - \theta^T D_t \theta - \theta^T \gamma_i [ \theta, X^i ] \bigr] \, ,
\ee
where $X_i$ are nine $N \times N$ Hermitean matrices, $D_t \equiv \partial_t - i [A(t), ..]$ is a covariant derivative operator containing a tenth $N \times N$ Hermitean matrix $A$,  the $\theta$ are sixteen fermionic superpartners which are spinors of $SO(9)$, and $g^2$ is a coupling constant of mass dimension $3$ (if the $X_i$ are taken to have canonical mass dimension of $1$). In the large $N$ limit, the `t Hooft coupling $\lambda \equiv g^2 N = g_s N l_s^{-3}$ (where $g_s$ and $l_s$ are string coupling constant and string length, respectively) is held fixed (see e.g. \cite{Ydri} for a review).

We will consider a high temperature state of this matrix model, and argue that, in the $N \rightarrow \infty$ limit, this state yields emergent space and emergent time (a time different from the time variable in the above action!). Based on results \cite{IKKT-SB} obtained in the context of the IKKT matrix model \cite{IKKT}, we argue that, as a function of the emergent time, the emergent space undergoes a phase transition during which the state breaks the spatial $SO(9)$ symmetry of the action spontaneously down to a $SO(3) \times SO(6)$ symmetry, with three spatial dimensions becoming large while the other six remain at the string scale. This is reminiscent of what happens in String Gas Cosmology \cite{BV} where the unwinding of string winding modes only allows three dimensions of space to become large while the others remain confined at the string scale.  We then compute the spectrum of both curvature perturbations and gravitational waves in this state, and demonstrate that they acquire spectra which are scale-invariant in the infrared \cite{us}. This result is again in good correspondence with what is obtained for curvature perturbations \cite{NBV} and gravitational waves \cite{BNPV} in the context of String Gas Cosmology  (see also \cite{SGCrevs} for a review).
 
These results suggest a new scenario for the early universe in the context of superstring theory: The initial state is a thermal state of the BFSS matrix model which yields emergent space and emergent time. Thermal fluctuations in this state yield scale-invariant spectra of energy density and pressure fluctuations. When the emergent phase ends with a transition to the usual Standard Cosmology phase of the expanding universe, the density and anisotropic pressure fluctuations induce curvature perturbations and gravitational waves, respectively. This suggests a scenario to obtain fluctuations consistent with current cosmological observations without requiring a phase of inflation (see also \cite{alternatives} for a review of other alternatives to inflation).

Since our starting point is a quantum mechanical model, the Big Bang singularity of an effective field theory description of cosmology is automatically resolved. One direct way to understand this is that the spacetime is given by the eigenvalue distribution of the bosonic matrices, and since these matrices are never trivial, there is no singularity in this model. 
In addition, there is no horizon problem: since it is space which emerges, the temperature is automatically uniform over the entire spatial section in a manifestly causal way.

In the high temperature limit, the fermions decouple and the BFSS model becomes equivalent to the bosonic part of the following IKKT \cite{IKKT} model:
\be \label{IKKTaction}
S \, = \,  -\frac{1}{g^2} {\rm{Tr}} \bigr( \frac{1}{4} [A^a, A^b] [A_a,A_b] + \frac{i}{2} \bar{\psi}_\alpha (C \Gamma^a)_{\alpha\beta} [A_a,\psi_\beta ]  \bigl) \,.
\ee
This is a pure matrix model (having neither background space nor background time), where $\psi_\alpha$ and $A_a$ ($\alpha =1,\ldots,16$, $a=0, \ldots, 9$) are $N\times N$ fermionic and bosonic Hermitian matrices, respectively, $\Gamma^{\alpha}$ are $D = 10$ gamma-matrices, and $C$ is the charge conjugation matrix\footnote{Note that the indices are contracted using the Minkowski $\eta_{a b}$ metric.}. The relationship is obtained by $A_i \equiv T^{-1/4} X_i^0$ , where $X_i^0$ are the zero modes of $X_i$ in a Fourier expansion of $X_i$ in the Euclidean time direction (recall that we are considering the matrix model at finite temperature).

The IKKT model has been studied extensively using numerical simulations (see e.g. \cite{IKKT-rev} for a review), and two key features emerge. Firstly, in the basis in which $A_0$ is diagonal, the spatial matrices are ``block-diagonal'' in the sense that the matrix elements decay as the distance from the diagonal increases. This is a highly nontrivial feature of the IKKT action, absent in typical matrix models. In the $N \rightarrow \infty$ limit, the eigenvalues of $A_0$ yield an emergent time. Without loss of generality we can order the eigenvalues such that time increases if we go down the diagonal. It is useful to define the time variable by averaging the $A_0$ eigenvalues over $n \ll N$ elements (see Figure 1). For $i = 1, \ldots, 9$, we can define the i'th spatial coordinate by
\begin{eqnarray}\label{Spatial_extent}
	R_i^2(n) :=  \lambda^{-4/3} \frac{1}{N\beta} \int_0^\beta dt \; {\rm{Tr}} ( {\tilde{X}}_i(t) )^2 \, ,
\end{eqnarray}
where $\tilde{X}_i(t)$ is the $n \times n$ spatial sub-matrix of $X_i$ centered at time $t$ (see Figure 1). The value of $n$ can be viewed as a comoving spatial coordinate, while ${\tilde{X}}_i$ gives the physical extent. As $n$ increases, we are hence probing larger and larger volumes of space. Importantly, $n$ is a fixed fraction of $N$, and hence the spatial extent defined by the above equation gives continuous space in the $N \rightarrow \infty$ limit \footnote{IR cutoffs are often introduced to make the emergent spatial sections finite, as required for numerical simulations.}. Note that the factor of $\lambda$ in the above equation ensures the correct mass dimensions.

\begin{figure}
    \includegraphics[scale = 0.5]{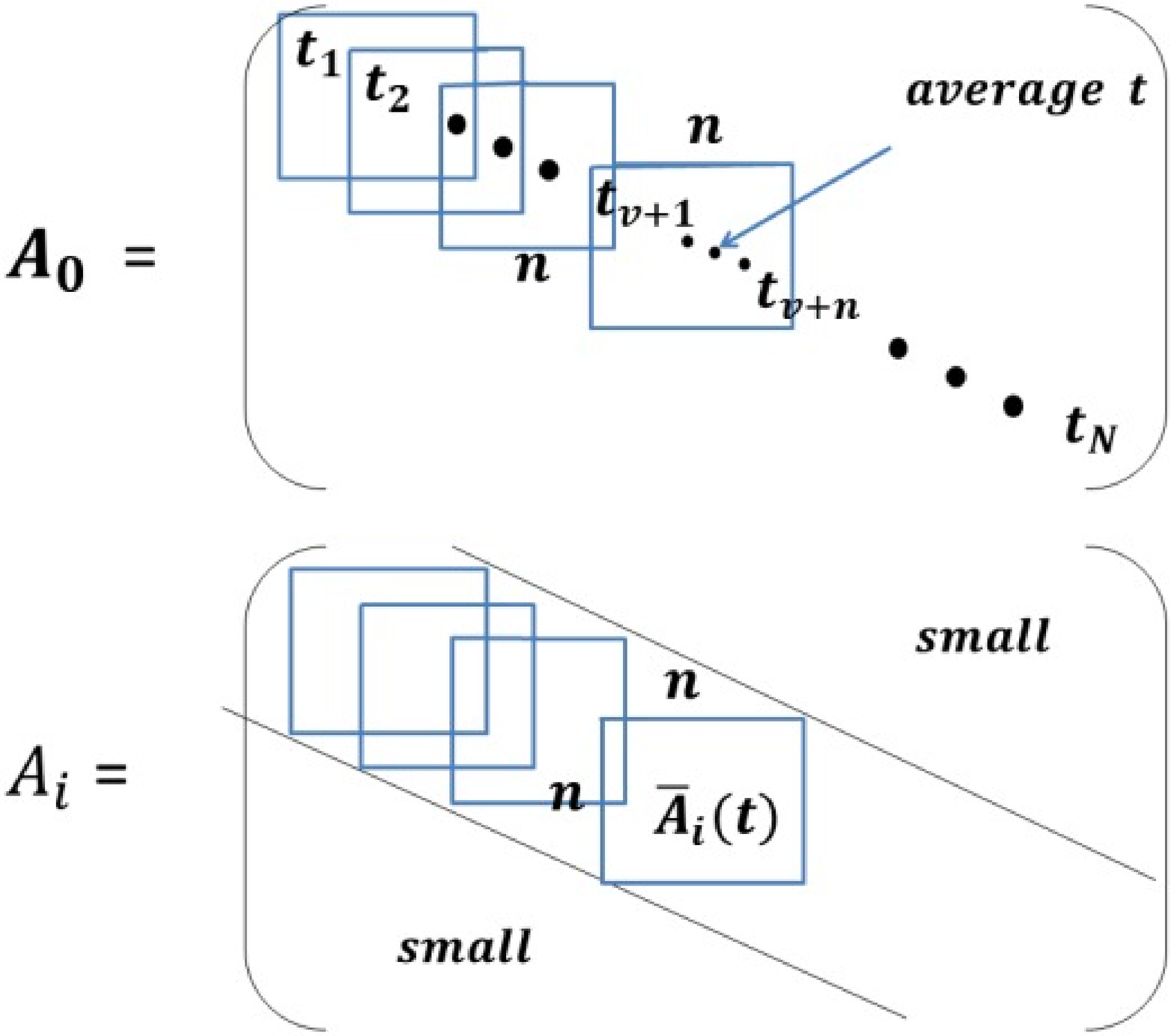}
    \caption{Structure of the matrices in the IKKT Model for large values of $N$. We work in the basis in which the temporal matrix (top panel) $A \equiv A_0$ is diagonal, and whose eigenvalues can be used to define emergent time as indicated. The spatial matrices $A_i$ (bottom panel) have ``block-diagonal form'' and can be used to define the sizes of the spatial dimensions. Note that, in the high temperature limit, the IKKT matrices $A_i$ are related to the zero modes $X_i^0$ of the BFSS matrices $X_i$ via the rescaling $A_i \equiv T^{-1/4} X_i^0$. This figure is taken from \cite{Ydri} with permission.} 
   \label{fig:conformal}
\end{figure}

The second conclusion of the numerical studies is that the spatial matrices show a phase transition: as time increases, three of the spatial dimensions grow while the other six remain constant at the string scale. This conclusion can also be obtained by free energy considerations \cite{IKKT-free}: starting from the matrix model partition function, we can perform a Gaussian expansion about configurations with various topologies, and one finds that the free energy is minimized for configurations which show the $SO(9) \rightarrow SO(3) \times SO(6)$ symmetry breaking. The presence of fermions plays an important role in obtaining this conclusion. Since the correspondence between the high temperature BFSS is only with the bosonic part of the IKKT model, we cannot directly apply the analysis of the IKKT studies. However, we are currently trying to apply similar analytical methods to the BFSS model and preliminary indications are that a similar $SO(9) \rightarrow SO(3) \times SO(6)$ symmetry breaking pattern is also applicable in the BFSS model \cite{us2}.

We have thus seen that a thermal state of the BFSS matrix model can generate an emergent space-time with three large spatial dimensions. Since the state is thermal (temperature $T$), there will be thermal fluctuations. Given the finite temperature partition function $Z(\beta)$ of the matrix model (where as usual $\beta$ is the inverse temperature), we can compute energy density and pressure fluctuations, as was done in the case of String Gas Cosmology in \cite{NBV, BNPV}. For example, the mean square energy density fluctuations in a box of radius $R$ are given by 
\be \label{eflucts}
\delta{\rho}^2(R) \, = \frac{T^2}{R^6} C_V(R)
\ee
where $C_V(R)$ is the specific heat capacity in a box of radius $R$, obtained by taking the partial derivative of the internal energy $E(\beta)$ with respect to temperature $T$ at constant volume. At late times, once a background metric emerges, the density fluctuations induce curvature perturbations (see e.g. \cite{MFB, RHBrev} for reviews of the theory of cosmological perturbations), whose power spectrum $P(k)$ on a momentum scale $k = 2\pi R^{-1}$ is given by
\be \label{spectrum}
P(k) \, = \, 16 \pi^2 G^2 (2 \pi)^{-6} k^2 T^2 C_V(R(k)) \, ,
\ee
where $G$ is Newton's gravitational constant. In a similar way, the amplitude of gravitational waves at late times is determined in terms of the off-diagonal pressure fluctuations.

The fluctuations emerging from the high temperature thermal state of the BFSS matrix model were computed in \cite{us}. The result for the specific heat capacity in a volume of radius $R$ in $d$ large dimensions was found to be
\be \label{specheat} 
C_V(R) \, = \, \frac{3 N^2}{4} \bigr( f_2 + f_1 \lambda^{4/3} T^{-2} R^2 \bigl) \, ,
\ee
with $f_1$ and $f_2$ being constants depending on $d$. Inserting this result into (\ref{spectrum}), we see that the first term yields a contribution to the power spectrum which is proportional to $T^2 R^{-2}$. This is a Poisson term which is expected for a thermal point particle process. This term is suppressed on infrared scales which are relevant to current cosmological observations. The second term in (\ref{specheat}), on the other hand, yields a scale-invariant contribution to the power spectrum which dominates on infrared scales. Making use of the scaling
\be
G^2 N^2 \lambda^{4/3} \, \sim \, \bigr( \frac{1}{l_s m_{pl}} \bigl)^4 \, ,
\ee
the power spectrum of curvature fluctuations in the infrared is independent of scale and has an amplitude
\be
{\mathcal{A}} \, \sim \, (l_s m_{pl})^{-4} \, .
\ee
If the string scale is of the order $10^{17} {\rm{GeV}}$, a value preferred for heterotic string phenomenology \cite{heterotic}, the amplitude of the fluctuations matches well with observations. 

Let us add a remark about the scale-invariance of the spectrum of fluctuations. The point is that at a fixed time $t$, we can consider a range of radii $R$ by taking a range of values of $n$. Thus, we are exploring the spectrum of perturbations over a range of box sizes. Now, if we vary the value of $n$ at a fixed location along the diagonal of the matrix $A$, we vary the number of diagonal elements which determine the time $t$ (see Figure 1). We expect that in the large $N$ limit the $n-$dependence of $t$ will be small. Nevertheless, taking this into account may introduce a small tilt to the spectrum.

The spectrum of gravitational waves is given by the off-diagonal pressure fluctuations. So far, we have only been able to compute the diagonal pressure fluctuations. These involve taking partial derivatives of the matrix theory partition function with respect to radius, while the energy density fluctuations involve a partial derivative with respect to temperature. The first term in (\ref{specheat}), the term which yields the Poisson contribution to curvature perturbations, evidently comes from a term in the internal energy which depends only on $T$ and not on $R$. Hence, that term in the internal energy does not contribute to the pressure fluctuations. The second term in the internal energy, on the other hand, has both $T$ and $R$ dependence and hence contributes to both energy density and pressure fluctuations. Hence, we find \cite{us} that the diagonal pressure fluctuations are given (up to a numerical factor) by the same expression as the term in the energy density fluctuations which dominates in the infrared. We can then conclude that the spectrum of gravitational waves will be scale-invariant (on all scales). Its amplitude is suppressed by a factor of $\alpha$ compared to the value of the curvature perturbations, which is the factor by which the off-diagonal pressure fluctuations are suppressed compared to the diagonal ones. It is an open question (and a question which is key to future observations) to determine the value of $\alpha$. We expect $\alpha$, which gives the predicted tensor-to-scalar ratio $r$, to be smaller than $1$, but not many orders of magnitude smaller. This expectation is supported by the concrete calculations in the case of String Gas Cosmology. To be consistent with current bounds on the amplitude of gravitational waves on cosmological scales, we require $\alpha \ll 10^{-1}$.

At first, from the point of view of the matrix model computations, the emergence of a scale-invariant spectrum of curvature fluctuations may come as a surprise. It is the result of the holographic scaling $C_V \sim R^2$ of the specific heat capacity. Such a scaling is expected in any holographic theory, and since holography is a crucial aspect of string theory, our result is hence less of a surprise. From the point of view of a thermal gas of fundamental strings, the high temperature scaling $C_V \sim R^2$ is obvious: at high temperatures, the gas of strings is dominated by winding modes whose thermodynamics is like that of point particles in one lower dimension, thus yielding the $C_V \sim R^2$ scaling. It would be of interest to explore in detail the physical origin of this scaling in the matrix model \footnote{In String Gas Cosmology, the emergent phase can be modelled by a quasi-static phase with zero pressure. In our analyses, the details of the non-geometric, emergent phase is yet to be worked out. However, in further support of a similar picture, the leading order terms in the free energy can be used to show that the background has an energy density although the pressure density is exactly equal to zero. The reason for this is the same as why there exists a scale-invariant part of the scalar spectrum, as well as a Poisson part, whereas for the gravitational wave spectrum, the spectrum only has a scale-invariant piece. In spite of our lack of understanding of the symmetry-breaking phase, it is interesting to note that, if modelled with an ideal fluid, the emergent phase for our model has the exact same characteristic as in String Gas Cosmology.}.

In conclusion, we have argued that the BFSS matrix model provides a new picture for the emergence of space, time and early universe cosmology. Starting from a thermal high temperature state of the BFSS model, the eigenvalues of the matrix $A$ yield a definition of an emergent time, a time coordinate which becomes continuous in the large $N$ limit. Working in the basis in which $A$ is diagonal, the nine spatial matrices yield yield a nine-dimensional space which also becomes continuous in the $N \rightarrow \infty$ limit. Local Lorentz invariance emerges since the matrix model contains the $\eta_{\mu \nu}$ matrix of Special Relativity (used to contract the indices in (\ref{BFSSaction})). As a function of time, the state of the system is argued to undergo a spontaneous symmetry breaking transition after which only three of the spatial dimensions become large. This gives hints of a dynamical mechanism for the compactification of the internal moduli space. Another heuristic way to understand the emergence of smooth spacetime in this picture is to consider the matrix model as a theory of D-branes. Roughly speaking, the eigenvalue distributions of the bosonic matrices denote the positions of D-branes. Only after the symmetry-breaking phase do we have a classical description of spacetime when the D-branes are well-separated and the matrices, corresponding to them, commute in this regime. On the other hand, the target space coordinates are inherently non-commutative at high temperatures when the off-diagonal elements cannot be ignored and denote string configurations connecting the D-branes.

We have computed the density and pressure perturbations in this thermal state and shown that they lead to scale-invariant spectra of curvature fluctuations and gravitational waves. Hence, a new picture of early universe cosmology emerges, a paradigm which is free of singularities, in which the Horizon Problem of Standard Big Bang cosmology is automatically solved, and which yields spectra of cosmological perturbations and gravitational waves consistent with current observations, without requiring a period of cosmological inflation. Note that the fluctuations are of thermal origin, and not of quantum vacuum origin, as they are in the case of inflationary cosmology \cite{ChibMukh}. 

Our scenario shares some common ideas with the recent proposal of \cite{Vafa2}. There, it was argued that the early universe phase is a topological phase which cannot be described in terms of a conventional effective field theory. The T-duality symmetry of string theory is a key aspect of the construction of \cite{Vafa2}, and thus the starting point is the same as in the case of String Gas Cosmology. In our setup, the topological phase is replaced by the emergent matrix model phase. Both our scenario and that of \cite{Vafa2} can be viewed as providing an underpinning for the ideas of String Gas Cosmology.

There are clearly lots of open issues. Some of the more technical challenges have been mentioned in the text, e.g. the challenge of demonstrating that the same symmetry breaking phase transition happens in the BFSS model as in the IKKT model. Possibly the most important open issue is to understand the emergence of the metric at late times, and the transition from the emergent phase to that of Standard Big Bang cosmology. In the case of String Gas Cosmology, this transition is obtained by the decay of string winding modes into radiation. It would be interesting to explore the physical details of the analog of this process in our matrix model cosmology.

Another important issue we have not addressed here is the ``Flatness Problem'', the need to explain the observed (near) flatness of spatial sections. We expect that similar arguments as those used in \cite{Vafa2} (or also more recently in \cite{Turok}) can be applied.  

{\tiny {\bf Acknowledgments:} We wish to thank B. Ydri for granting us permission to use Fig. \ref{fig:conformal} from \cite{Ydri}. The research at McGill is supported in part by funds from NSERC and from the Canada Research Chair program. SB is supported in part by the Higgs Fellowship in Theoretical Physics.}

\end{document}